\documentstyle[10pt]{article}
\begin{document}
\newcommand{\Si}{\Sigma}
\newcommand{\tr}{{\rm tr}}
\newcommand{\ad}{{\rm ad}}
\newcommand{\Ad}{{\rm Ad}}
\newcommand{\ti}[1]{\tilde{#1}}
\newcommand{\om}{\omega}
\newcommand{\Om}{\Omega}
\newcommand{\de}{\delta}
\newcommand{\al}{\alpha}
\newcommand{\te}{\theta}
\newcommand{\vth}{\vartheta}
\newcommand{\be}{\beta}
\newcommand{\la}{\lambda}
\newcommand{\La}{\Lambda}
\newcommand{\D}{\Delta}
\newcommand{\ve}{\varepsilon}
\newcommand{\ep}{\epsilon}
\newcommand{\vf}{\varphi}
\newcommand{\G}{\Gamma}
\newcommand{\ka}{\kappa}
\newcommand{\ip}{\upsilon}
\newcommand{\Ip}{\Upsilon}
\newcommand{\ga}{\gamma}
\newcommand{\ze}{\zeta}
\newcommand{\si}{\sigma}
\def\vka{\varkappa}
\def\bfa{{\bf a}}
\def\bfb{{\bf b}}
\def\bfc{{\bf c}}
\def\bfd{{\bf d}}
\def\bfm{{\bf m}}
\def\bfn{{\bf n}}
\def\bfp{{\bf p}}
\def\bfu{{\bf u}}
\def\bfv{{\bf v}}
\def\bft{{\bf t}}
\def\bfx{{\bf x}}
\newcommand{\li}{\lim_{n\rightarrow \infty}}
\newcommand{\mat}[4]{\left(\begin{array}{cc}{#1}&{#2}\\{#3}&{#4}
\end{array}\right)}
\newcommand{\mathr}[9]{\left(\begin{array}{ccc}{#1}&{#2}&{#3}
\\{#4}&{#5}&{#6}\\
{#7}&{#8}&{#9}
\end{array}\right)}
\newcommand{\beq}[1]{\begin{equation}\label{#1}}
\newcommand{\eq}{\end{equation}}
\newcommand{\beqn}[1]{\begin{eqnarray}\label{#1}}
\newcommand{\eqn}{\end{eqnarray}}
\newcommand{\p}{\partial}
\newcommand{\di}{{\rm diag}}
\newcommand{\oh}{\frac{1}{2}}
\newcommand{\su}{{\bf su_2}}
\newcommand{\uo}{{\bf u_1}}
\newcommand{\GL}[1]{{\rm GL}({#1},{\bf C})}
\newcommand{\SL}[1]{{\rm SL}({#1},{\bf C})}
\newcommand{\sll}[1]{{\rm sl}({#1},{\bf C})}
\def\sln{{\rm sl}(N,{\bf C})}
\newcommand{\gl}{gl(N,{\bf C})}
\newcommand{\PSL}{{\rm PSL}_2({\bf Z})}
\def\f1#1{\frac{1}{#1}}
\newcommand{\rar}{\rightarrow}
\newcommand{\upar}{\uparrow}
\newcommand{\sm}{\setminus}
\newcommand{\ms}{\mapsto}
\newcommand{\bp}{\bar{\partial}}
\newcommand{\bz}{\bar{z}}
\newcommand{\bA}{\bar{A}}
\newcommand{\begarl}{\begin{array}{l}}
\newcommand{\enarr}{\end{array}}
\newcommand{\begarc}{\begin{array}{c}}
\newcommand{\sect}[1]{\setcounter{equation}{0}\section{#1}}
\renewcommand{\theequation}{\thesection.\arabic{equation}}
\newtheorem{predl}{Proposition}[section]
\newtheorem{defi}{Definition}[section]
\newtheorem{rem}{Remark}[section]
\newtheorem{cor}{Corollary}[section]
\newtheorem{lem}{Lemma}[section]
\newtheorem{theor}{Theorem}[section]

\vspace{0.3in}
\begin{flushright}
 ITEP-TH-18/99\\
\end{flushright}
\vspace{10mm}
\begin{center}

{\Large\bf
Non-autonomous Hamiltonian systems related to highest Hitchin integrals}
\footnote
{Contribution in the Proceedings 
"International Seminar on Integrable systems". In memoriam Mikail V. Saveliev.
Bonn, February, 1999}
\vspace{5mm}

A.M.Levin\\
{\sf Max-Plank-Institut f\"{u}r Matematik, Bonn}\\
{\sf Institute of Oceanology, Moscow, Russia,} \\
{\em e-mail andrl@landau.ac.ru}\\
M.A.Olshanetsky
\\
{\sf Max-Plank-Institut f\"{u}r Matematik, Bonn}\\
{\sf Institute of Theoretical and Experimental Physics, Moscow, Russia,}\\
{\em e-mail olshanet@heron.itep.ru}\\

\vspace{5mm}
\end{center}

\begin{abstract}
We describe non-autonomous Hamiltonian systems coming from the Hitchin 
integrable systems. The Hitchin integrals of motion depend on
the ${\cal W}$-structures  of the basic curve. The parameters of the 
${\cal W}$-structures play the role of times. In particular, the
quadratic integrals dependent on the complex structure (${\cal W}_2$-structure) 
of the basic curve and times are coordinate on the Teichm\"{u}ller space. 
The corresponding flows are the monodromy preserving equations such as 
the Schlesinger equations, the Painlev\'{e} VI equation and their 
generalizations. The equations corresponding to the highest integrals are
 monodromy preserving conditions with respect to changing of the 
${\cal W}_k$-structures ($k>2$). They are derived by the symplectic reduction 
from the gauge field theory on the basic curve interacting with 
${\cal W}_k$-gravity. As by product we obtain the classical Ward identities 
in this theory. 
\end{abstract}
\vspace{0.3in}
\begin{flushright}
{\it In memory of Mikhail Saveliev}\\
\end{flushright}

\vspace{0.12in}
\bigskip

\section {Introduction}
\setcounter{equation}{0}

Infinite-dimensional symmetries corresponding to the ${\cal W}$-algebras 
play a central role in two-dimensional physics (see \cite{Hu} for a review).
Here we investigate classical Hamiltonian systems incorporated in a gauge
theory on a Riemann curve interacting with the ${\cal W}$ gravity.
Whereas the ${\cal W}_2$-gravity has a natural geometric background, there is
no satisfactory geometric understanding of the origin of the 
${\cal W}_k,~(k>2)$ theory. There exists a viewpoint on the 
${\cal W}_k$-gravity as the geometry of certain two-dimensional surfaces 
embedded in $k$-dimensional affine space. One of Misha Saveliev's works
\cite{Sa} is an important step in this direction. Here we analyze
interrelations between ${\cal W}$-geometry and integrable systems.
On the later subject Misha Saveliev has had a significant influence.

We starting from some subclass of classical completely integrable system
with phase flows having the Lax form
\beq{1.1}
\p_sL=[L,M_s],~~\p_s=\frac{\p}{\p t_s}.
\eq
We assume that \\
i)$L$ takes value in a complex Lie algebra. We restrict ourself to the 
case $\sln$;\\
ii)$L=L(z)$, where $z$ is a spectral parameter lying on the Riemann
curve $\Si_{g,n}$ of genus $g$ with $n$ marked points. $L$ can have first
order poles in the marked points. Thereby, we exclude in what follows the
Stokes phenomena
\footnote{This restriction was partly resolved in
 the recent paper \cite{U}.}.
The integrable systems on curves without marked points was considered by 
Hitchin \cite{H1}.
In this case nontrivial equations arise only on the high genus curves $(g>1)$. 
Up to the recent time there were no explicit examples of such type
integrable systems\footnote{See, however, \cite{GP,Ga} for 
$g=2,~L\in sl(2,{\bf C})$ }.
The generalization of Hitchin approach to the matrices with first order poles
in the marked points \cite{Ne} allowed to include in this scheme some well 
known completely integrable models like the Toda and the Calogero-Moser systems.
The quantum counterpart of these systems are the Knizhnik-Zamolodchikov-Bernard
equations for conformal blocks in the WZW theory on the critical level, 
while the original Hitchin systems correspond to equations for partition functions.

In the Hitchin type systems $\tr (L^k)=<L^k>$ being integrated over $\Si_{g,n}$
are integrals of motion. Before the integration one
should take into account that $L$ are $(1,0)$-forms on
$\Si_{g,n}$ in some fixed complex structure. To integrate $<L^k>$ one
should multiply it on $(-k+1,1)$-differentials  
$\rho_{k,s}=\rho_{k,s}(z,\bz)\p_z^{k-1}\otimes d\bz$. The index $s$ arises
in the following way. The  operator
$\rho_{k,s}(z,\bz)\p_z^{k-1}$ is defined in local coordinates of the point
$(z,\bz)$. The fields $\rho_{k,s}(z,\bz)\p_z^{k-1}$ that can be represented as 
$\bp$ derivative do not contribute in the Hamiltonians.
In this way $\rho_{k,s}$ can be chosen from $H^1(\Si_g, \G^{k-1})$,
where $\G$ are vector fields on $\Si_{g,n}$ vanishing in the marked points.
 Then $s$ enumerates the basis in $ H^1(\Si_g, \G^{k-1})$.

Represent the differentials as
\beq{1.3}
\rho_{k,s}=t_{k,s}\rho_{s,k}^0,
\eq
where $\{\rho_{s,k}^0\}$ is a fixed basis in $ H^1(\Si_{g,n}, \G^{k-1})$.
The integrals of motion take the form
\beq{1.1b}
H_{k,s}=\f1{k}\int_{\Si_g}<L^k>\rho_{k,s}^0,~~
(k=1,\ldots,N,~s=1,\ldots).
\eq
The important class of Hamiltonian equations occurs when  $t_{k,s}$
are considered as "times". In this case $H_{k,s}$ become the Hamiltonians of 
non-autonomous systems, where times are
related to the deformations of the internal structure of the base
curve $\Si_{g,n}$.
  It turns out that the phase flows of these systems are described by the 
following modification of the Lax equation (\ref{1.1})
\beq{1.1a}
\p_aL-\p M_a+[M_a,L]=0,~~a=(k,s), \p=\frac{\p}{\p z}.
\eq
Consider first the quadratic Hamiltonians $H_{2,s}$. In this case
$\rho_{2,s}=\mu_s$ are the $(-1,1)$-differentials 
(the Beltrami differentials).
 They are $(0,1)$-forms taking values in the vector fields on $\Si_{g,n}$. 
Here we deal with the Lie algebra of vector fields and
the group of local diffeomorphisms of $\Si_{g,n}$. Roughly speaking the space
$ H^1(\Si_{g,n}, \G)$ can be defined as the space of smooth 
$(-1,1)$-differentials on  $\Si_{g,n}$ modulo global diffeomorphisms action.
The elements from $ H^1(\Si_{g,n}, \G)$ play role of deformation 
parameters of the complex structure on $\Si_{g,n}$.
 In the genus zero case  the complex structure is defined by the positions of 
marked points $t_s=x_s-x_s^0$ and (\ref{1.1a}) leads to
 the Schlesinger equation. Another interesting examples including a particular 
family of the
Painlev\'{e} VI equation occur when the basic curve
is an elliptic curve with marked points \cite{LO}. 

For $k>2$ the differentials $\rho_{k,s}$ do not
generate a Lie algebra. Due to this fact they have not  natural geometric
description. Along with the dual objects (opers \cite{BD}) they generate the 
so called ${\cal W}_k$-geometry of the basic curve $\Si_{g,n}$ \cite{F,GLM}. 
${\cal W}_k$-geometry is a generalization of ${\cal W}_2$-geometry 
that coincides with the space of projective structures of $\Si_{g,n}$. 

The invariant object associated with the Lax form (\ref{1.1})
of integrable systems is the spectral curve ${\cal C}$
$$
{\cal C}: ~f(\la,z):=\det(\la+L(z))=0.
$$
It is $N$ cover of the basic curve $\Si_{g,n}$. Then the Prym variety
$Prym({\cal C}/\Si_g)$ 
is the Liouvillian torus of the completely integrable system (\ref{1.1})
\cite{H1}.
There are different parametrizations of ${\cal C}$. The two standard
parametrizations are the set
of the Hamiltonians $H_{k,s}$ (\ref{1.1b}) and the action variables.
 The differentials $\rho_{k,s}$
(\ref{1.3}) provide another parametrization of ${\cal C}$ related to the 
${\cal W}_N$ structure of the basic curve. For small times (\ref{1.1a})
describes an evolution of ${\cal C}_t$ near the fixed curve ${\cal C}_0$ .

The main goal of this paper is to investigate the dynamical systems 
(\ref{1.1a}) associated 
with the ${\cal W}_k$-geometry. There are two important aspects of this
investigation. First, the quantum analog of these systems are higher order
Knizhnik-Zamolodchikov-Bernard equations beyond the critical level.
 It means that conformal
blocks in the WZW theory satisfy some analogue of nonstationar
Schr\'{o}dinger equations with higher order Casimirs and times 
$t_{k,s},k>2$. We don't aware of some explicit examples of these type of 
equations. On the other hand, the investigation of the higher 
${\cal W}$-geometries
is interesting by itself. It is by no means an easy problem because apparently 
these geometries are not of the Klein type - there are no evident group
symmetries related to them. The connection of the ${\cal W}$-geometries with the 
integrable systems opens a new way for investigations of the 
${\cal W}$-geometries. Namely, one can apply tools developed for integrable 
systems, such as the 
Whitham quantization method \cite{Kr1} which is well adjusted to 
the analysis of the perturbations of integrals of motion
(\ref{1.1b}). 

Our construction is based on the approach developed in \cite{F}.
The ${\cal W}_k$ structures are described there as a result of Hamiltonian 
reduction with respect to a maximal parabolic subgroups of $\SL{k}$.
We generalize this approach in two directions:\\
$\bullet$ we consider Riemann curves with marked points
adding some additional data in the marked points;\\
$\bullet$ in addition to the ${\cal W}$ fields we include gauge fields.\\
As a result we obtain the classical Ward identities for ${\cal W}$-gravity
interacting with gauge theory on Riemann curves with marked points.
To obtain monodromy preserving flows we exclude the half
of ${\cal W}$ fields leaving only differentials $\rho_{k,s}$, which play 
role of "times".
This procedure is also based on the Hamiltonian reduction with respect
to symmetries generated by the Sugawara type constraints.
On this stage the flows are rather trivial, since the systems are free.
Finally, the Hamiltonian reduction based on the gauge symmetries leads
to nontrivial dynamical systems.

The plan of the paper is follows. In the first sections we revise
the derivation of the equations preserving monodromies using the
projective structure of basic curves.
The similar program is done further in details for the ${\cal W}_3$ case. 
In conclusion we discuss shortly the general case.

\section{Projective structures on Riemann curves and symplectic geometry}
\setcounter{equation}{0}

{\bf 1. Projective structures.}\\
Let us fix the complex structure on  $\Si_{g,n}$ by choosing a pair
of local coordinates $(z,\bz)$ and the corresponding operators
$(\p,\bp)$.
The deformed complex structure
can be read off from the solutions of the Beltrami equation
\beq{3.0}
(\bp+\mu\p)F=0.
\eq
Locally,  $F(z,\bz)$ is the diffeomorphism
$$
w=F(z,\bz),~\bar{w}=\bar{F}(z,\bz)
$$
and  the Beltrami differential is
$$
\mu(z,\bz)=-\frac{\bp\bar{F}}{\p F}.
$$
It defines the new complex structure operator $\p_{\bar{w}}:=\bp+\mu\p$.
 Two Beltrami differentials produce equivalent complex structures if they are
related by a global holomorphic diffeomorphisms of $\Si_{g,n}$
\beq{hd}
w=z-\ep(z,\bz),~\bar {w}=\bz.
\eq
We assume that the Lie algebra ${\cal V}_{g,n}$ of corresponding vector fields 
on  $\Si_{g,n}$ is specified by their behavior near the marked points
\beq{3.01}
{\cal V}_{g,n}=\{\ep(z,\bz)\p~|~\ep(z,\bz)=O(z-x_a)\}
\eq
Under the holomorphic diffeomorphisms $\mu$ transforms as $(-1,1)$-differential.
The vector fields act on $\mu$ as
\beq{3.00}
j_{\ep}\mu=-\ep\p\mu +\mu\p\ep+\bp\ep.
\eq

We specify the dependence of $\mu$ on the positions of
the marked points in the following  way.
Let ${\cal U}'_a\supset{\cal U}_a$ be two vicinities
 of the marked point $x_a$
such that ${\cal U}'_a\cap{\cal U}'_b=\emptyset$ for $a\neq b$.
Let $\chi_a(z,\bz)$ be a smooth function
\beq{cf}
\chi_a(z,\bz)=\left\{
\begin{array}{cl}
1,&\mbox{$z\in{\cal U}_a$ }\\
0,&\mbox{$z\in\Si_g\setminus {\cal U}'_a.$}
\end{array}
\right.
\eq
Introduce times related to the positions of the
marked points $t_{2,a}=x_a-x_a^0$.
Then $\mu$ can be represented as
\beq{6.1}
 \mu=\sum_{a=1}^nt_{2,a}\mu^0_a,~~
\mu^0_a=\bp \ep_a(z,\bz),~~\ep_a(z,\bz)=\chi_a(z,\bz),~~(t_{2,a}=x_a-x_a^0).
\eq

Let $T$ be the
projective connection on $\Si_{g,n}$, i.e. $T$ is transformed under 
the holomorphic diffeomorphisms as
$(2,0)$-differential up to the addition of the Schwarzian derivative.
Locally it means that
\beq{3.1b}
j_{\ep}T(z,\bz)=-\ep\p T-2T\p\ep-\frac{\ka^2}{2}\p^3\ep.
\eq
Here $\ka$ is a parameter, which later will play role of the "Planck constant" 
in the Whitham quantization.
We assume that $T$ has poles at the marked points $x_a,(a=1,\ldots,n)$
  up to the second order:
\beq{3.1a}
T|_{z\rar x_a}\sim\frac{T_{-2,a}}{(z-x_a)^2}+\frac{T_{-1,a}}{(z-x_a)}+\ldots
\eq
Let  $\ti{\cal W}_2$ be the space of the pairs $(T,\mu)$ on $\Si_{g,n}$ with the
behavior near the marked points defined by (\ref{6.1}),(\ref{3.1a}).

\begin{defi}
The space ${\cal W}_2$ of projective structure on $\Si_{g,n}$ is the subset
of $\ti{\cal W}_2$ that satisfies the equation
\beq{3.1}
(\bp+\mu\p+2\p\mu)T=\oh\p^3\mu,
\eq
 with fixed values of 
${\bf T}_2=(T_{-2,1},\ldots,T_{-2,l})$ in (\ref{3.1a}).
\end{defi}

Let $\psi$ be a $(-\oh,0)$ differential. Then (\ref{3.1}) 
is the compatibility condition for the linear system
\beq{3.2}
(\bp+\mu\p -\oh\p\mu) \psi=0,
\eq
\beq{3.3}
(\p^2-T)\psi=0.
\eq

Consider  two linear independent solutions $\psi_1,\psi_2$ to the
system.  The projective structure $(T,\mu)$ can be equivalently defined
by their ratios $F=\psi_1/\psi_2$. In fact,  it follows from (\ref{3.2}) that
$F$ satisfies the Beltrami equation (\ref{3.1}). Therefore, $\mu=\bp F/\p F$.
On the other hand, from (\ref{3.3}) $\psi_1=(\p F)^\oh$ and $T={\cal S}_z(F)$,
where ${\cal S}_z(F)$ is the Schwarzian derivative of $F$. Arbitrary
linear independent solutions of the system (\ref{3.2}),(\ref{3.3}) are
obtained from $\psi_1,\psi_2$ by the $\SL{2}$ transform. It results in the
M\"{o}bius transform of $F$ and does not
change $T={\cal S}_z(F)$. Thus, the relations of any two independent solutions
defines projective structure as well.

\bigskip
\noindent

{\bf 2. Symplectic reduction with respect to diffeomorphisms.}\\
The space ${\cal W}_2$ is similar to the space of flat connections on $\Si_{g,n}$
for some gauge group. It is a symplectic manifold,
which can be derived from the affine space of smooth connections via
the symplectic reduction. The flatness condition plays the role 
of the moment constraint equation. The similar procedure can be applied to
the space $\ti{\cal W}_2$ to obtain ${\cal W}_2$. In this case the gauge group is 
replaced by the group
of holomorphic diffeomorphisms of $\Si_{g,n}$ (\ref{hd}). The space 
$\ti{\cal W}_2$ can be endowed with the symplectic structure
\beq{3.4}
\om=-\ka^{-1}\int_{\Si_g}\de T\de\mu.
\eq
The action of the vector fields ${\cal V}_{g,n}$ (\ref{3.00}),(\ref{3.1b})
is the symmetry of $\om$. The Hamiltonian of this action
$$
H_\ep=-\ka^{-1}\int_{\Si_g}\ep
[(\bp+\mu\p+2\p\mu)T-\frac{\ka^2}{2}\p^3\mu]
$$
produces the moment map 
\beq{3.04}
m:\ti{W}_2\rar {\cal V}^*_{g,n},~~m=(\bp+\mu\p+2\p\mu)T-\frac{\ka^2}{2}\p^3\mu,
\eq
where ${\cal V}^*_{g,n}$ is the dual space to the algebra ${\cal V}_{g,n}$  
of vector fields.
It is the space of $(2,1)$-forms on $\Si_{g,n}$. As it follows from
(\ref{3.01}) in the neighborhoods of marked points elements $y\in{\cal V}^*_{g,n}$
take the form
\beq{3.03}
y\sim b_{1,a}\p\de(x_a)+b_{2,a}\p^2\de(x_a)+\ldots.
\eq 
Thereby, the terms $T_{-1,a}\de(x_a)$ that arise in $\bp T$ (\ref{3.04})
from the first order 
poles of $T$ (\ref{3.1a}) are projected out from the moment map $m$ (\ref{3.04}).
We take
$$
m=-\sum_{a=1}^nT_{-2,a}\p\de(x_a).
$$
It follows from (\ref{3.1a}),(\ref{3.04}) and (\ref{3.03}) that we put in (\ref{3.03}) 
$ b_{1,a}=-T_{-2,a},~~b_{k,a}=0,~k>1.$
Thus, the condition (\ref{3.1}) that distinguish ${\cal W}_2$ in $\ti{\cal W}_2$
is the moment constraint with respect to the action of the diffeomorphisms.
\bigskip
\noindent

{\bf 3."Drinfeld-Sokolov" approach.}
In \cite{F} another procedure was proposed, which resembles
the Drinfeld-Sokolov approach.
Shortly, it looks as follows. Consider the affine space ${\cal N}_2$ of 
$\SL2$ smooth flat connections on $\Si_{g,n}$
$$
{\cal N}_2=\{adz+\bar{a}d\bz\},
$$
\beq{3.7}
F(a,\bar{a})=\bp a-\p\bar{a}+[a,\bar{a}]=0.
\eq
The field $a$ has  poles in the marked points  up to the second order.
The space ${\cal N}_2$ has the standard symplectic form
\beq{3.5}
\om'=\int_{\Si_g}\tr(\de a\de\bar{a}).
\eq
The form is invariant under the gauge transform
$$
a\rar g^{-1}ag+g^{-1}\p g,~~\bar{a}\rar g^{-1}\bar{a}g+g^{-1}\bp g,~~
g\in{\cal G}={\rm Map}(\Si_{g,n},\SL2)
$$
We assume that the Lie algebra of the gauge group Lie$({\cal G})$ is
specialized by the behavior of its matrix element $x^{12}$ near the
marked points
\beq{3.5a}
x^{12}|_{z\rar x_a}=O(z-x_a),~~(x^{JK})\in{\rm Lie}({\cal G}).
\eq
The flatness (\ref{3.7}) is the moment constraint with respect to the action 
of the gauge group.
The form $\om'$ is degenerated - it vanishes on the orbits 
of the gauge
group, because we only put the moment condition (\ref{3.7}) and do not
fix the gauge. If we do it we come to the finite-dimensional space of flat
connections, but it is necessary to leave two fields $T$ and $\mu$.
The trick proposed in \cite{F} is to fix the gauge  with respect to
 the Borel subgroup $B$ of the lower  triangular matrices.
It was proved there that it allows to obtain from the  space ${\cal N}_2$
the  space of projective  connections ${\cal W}_2$.

The gauge freedom allows to fix a generic matrix $a$ in the form
\beq{3.6}
a=\mat{0}{1}{T}{0},
\eq
where $T$ is a new field satisfying (\ref{3.1a}). The first order poles of
$T$ do not contribute in the moment equation (\ref{3.7}) since they are eaten
by the gauge transform (see (\ref{3.5a})).
The moment equation (\ref{3.7})
allows to express all matrix elements of $\bar{a}$ in terms of $T$
 and new field $\mu$
\beq{3.8}
\bar{a}=\mat{\oh\p\mu}{-\ka^{-1}\mu}{-\ka^{-1}\mu T+\oh\ka\p^2\mu}{-\oh\p\mu}.
\eq
The flatness condition becomes trivial for all matrix element except  
$F_{(2,1)}$. It can be checked that it just coincides with the projectivity
condition (\ref{3.1}). 

The linear system for $(-\oh,0)$ differentials
 $\psi$
\beq{3.9a}
(\bp+\bar{a})
\left(\begin{array}{c}
\psi\\
\p \psi
\end{array}\right)=0,
~~(\ka\p+a)
\left(\begin{array}{c}
\psi\\
\p \psi
\end{array}\right)=0
\eq
is consistent due to (\ref{3.7}). It is the matrix form
of (\ref{3.2}),(\ref{3.3}) for the special form
of $a$ and $\bar{a}$ (\ref{3.6}),(\ref{3.8}).

The original symplectic structure $\om'$
(\ref{3.5}) is reduced to $\om$ (\ref{3.4}) on the space of the projective
structures ${\cal W}_2$.

Though, the diffeomorphisms do not arise in this approach, they are hidden
in this construction. To demonstrate it, calculate the commutator
of two matrices $\bar{a}_1$,  $\bar{a}_2$
$$
[\bar{a}_1(\mu_1,T_1),\bar{a}_2(\mu_2,T_2)]_{(1,2)}=
\mu_1\p\mu_2-\mu_2\p\mu_1.
$$
Thus, the commutator of matrix $\bar{a}$ reproduces the commutator of
vector fields.

\section {Isomonodromic deformations and projective structures}
\setcounter{equation}{0}
{\bf 1.${\cal W}_2^N$-structures - definition.}\\
Consider some projective structure on $\Si_g$ defined by the linear system
(\ref{3.2}),(\ref{3.3}). We generalize it in the following way.
Consider the vector $\SL{N}$-bundle $V$
over $\Si_g$.
Let $(A,\bA)$ be connections in $V$ corresponding to the complex structure 
we have fixed.\\
We assume that\\
$\bullet$ the connection $\bA$ is smooth;\\
$\bullet$  the connection $A$ has first order poles in the marked points
\beq{4.0}
A\sim\frac{A_{-1,a}}{z-x_a}+A_{0,a}+\ldots,~a=1,\ldots,n.
\eq
In addition to this data consider the set of coadjoint orbits of  $\SL{N}$
in the marked points
\beq{4.00}
({\cal O}_1,\ldots,{\cal O}_n),~~
{\cal O}_a=\{(p_a=g_ap_a^0g_a^{-1}|g_a\in\SL{N}\}.
\eq
Here $p_a^0$ specifies the choice of the orbit ${\cal O}_a$.
To reconcile this data with the projective structures instead of the operator
(\ref{3.3})
we consider in what follows the matrix operator 
$$
(\ka\p+A)^2-T.
$$
Define the matrices
\beq{4.3}
\ti{T}=T-A^2-\ka\p A.
\eq
and
\beq{f11}
f_1=-\bA+\oh\p\mu{\bf 1}_N-\f1{\ka}\mu A.
\eq
We skip the multiplication on the scalar matrices ${\bf 1}_N$ in what follows.

Define the space $\ti{\cal W}_2^N$ of fields 
$(T,\mu),(A,\bA),{\bf p}=(p_1,\ldots,p_l)$
on $\Si_{g,n}$, where the behavior of $\mu,T$ and $A$  in the neighborhood 
of marked points satisfies (\ref{6.1}),(\ref{3.1a}) and (\ref{4.0}) 
correspondingly.

\begin{defi}
${\cal W}_2^N$-structure on $\Si_g$ is the subset of  $\ti{\cal W}_2^N$
satisfying  the following identities:
\beq{4.5}
(\bp+\mu\p+2\p\mu)\ti{T}-\oh\ka^2\p^3\mu=[\ti{T},\bA+\f1{\ka}\mu A]
+2\ka A\p f_1 ,
\eq
\beq{4.6}
\bp A-\ka\p\bA+[A,\bA]=0, 
\eq
\beq{4.6a}
A_{-1,a}=-p_a,~(a=1,\ldots,n),
\eq
\beq{4.6b}
T_{-2,a}=\frac{<(p_a^0)^2>}{N}
~(a=1,\ldots,n).
\eq
\end{defi}
In the first relation the right hand side is the contribution of the
gauge fields in the classical Ward identity for the gravitational fields
$(T,\mu)$. The second identity is the standard flatness condition for
the gauge fields. The last identity expresses the most singular terms 
$T_{-2,a}$ of the projective connection $T$ in terms of the second Casimirs
parametrizing the orbits ${\cal O}_a$.

There is a straightforward generalization of the linear system
(\ref{3.2}),(\ref{3.3}) defining the projective structure.
Let $\Psi$ be the element of the space of sections
$\Om^{(-\oh,0)}(\Si_g,{\rm Aut}V)$.
\begin{predl}
The equations (\ref{4.5}),(\ref{4.6}),
 are the compatibility conditions for the linear system
\beq{4.1}
(\bp+\mu\p -\oh\p\mu+\bA+\f1{\ka}\mu A) \Psi=0,
\eq
\beq{4.2}
[(\ka\p+A)^2-T]\Psi=0.
\eq
\end{predl}

The space ${\cal W}_2^N$ can be endowed with the degenerated symplectic 
 form
\beq{4.7}
\om=\int_{\Si_{g,n}}[-\frac{N}{\ka}\de T\de\mu+
2<\de A,\de\bA>]+4\pi i\sum_{a=1}^n\de <p_a,g_a^{-1}\de g_a>.
\eq
The last sum in (\ref{4.7}) is the contribution of symplectic forms on 
the coadjoint orbits. We demonstrate below that this form is natural and come
from a Hamiltonian reduction procedure.

\bigskip
\noindent
{\bf 2. Symplectic construction.}\\
Consider the vector bundle $V_{2N}=V_2\otimes V_N$
over $\Si_g$
with the structure group\\
$G=S(\GL{2}\otimes\GL{N})\sim \SL{2N^2}$. In the
nondeformed complex structure on $\Si_{g,n}$ the connection operators are
$$
\ka\p-{\cal A},~~\bp-\bar{\cal A},
$$
where ${\cal A},\bar{\cal A}$ take values in Lie$(S(\GL{2}\otimes\GL{N}))$.
The first components of ${\cal A}$ and $\bar{\cal A}$ act on the sections of 
$1$-jets of $\Om^{(-\oh,0)}(\Si_{g,n})$, while the second components act
on the sections of the bundle $\Om^{(-\oh,0)}(\Si_{g,n},{\rm Aut}V)$.

Define the space  ${\cal K}_2$ as the space of connections 
$({\cal A},\bar{\cal A})$     
with coadjoint orbits attached in the marked points 
$$
({\cal A},\bar{\cal A});
({\bf 1}_2\otimes{\bf p})=({\bf 1}_2\otimes p_1,\ldots,{\bf 1}_2\otimes p_l)
$$
 with the following additional restrictions:\\
$\bullet$ the ${\cal A}$ component are gauge equivalent
to the special form
\beq{4.60a}
{\cal A}\sim DS\otimes{\bf 1}_N+{\bf 1}_2\otimes(-A),
\eq
where
$$
DS=\mat{0}{1}{T}{0};
$$ 
$\bullet$ near the marked points $T$ has the form (\ref{3.1a});\\
$\bullet$ near the marked points $A$ has the form (\ref{4.00}).

Propositions 3.1 follow from the following statement
\begin{predl}
 ${\cal W}_2^N$ is the subset ${\cal K}_2$ satisfying the flatness condition 
\beq{4.50}
[\ka\p-{\cal A},\bp-\bar{\cal A}]=0.
\eq
\end{predl}
{\sl Proof.}\\
The space ${\cal K}_2$ has the standard symplectic form  
\beq{4.30}
\om'=\int_{\Si_{g,n}}<<\de{\cal A},\de\bar{\cal A}>>+
4\pi i\sum_{a=1}^n\de <p_a,g_a^{-1}\de g_a>
\eq
Here $<<,>>$ is the trace in the tensor product, and $<,>$ is the trace
in the $V_N$ space. 

Introduce the following group of gauge transforms. It is the smooth maps
$$
{\cal G}=\{{\rm Map}(\Si_{g,n}, \SL{2N^2})\}
$$
with additional restrictions for the maps near the marked points. We formulate
them on the Lie algebra Lie$({\cal G})$ level. Let $x^{IJ}_{\al\be}$ be 
the matrix element 
in Lie$({\cal G})$. Here the upper indices $I,J$ are related to the space $V_2$ 
and the lower indices $\al,\be$ to the space $V_N$. We assume that in the 
neighborhood of the marked point
\beq{ge2}
x^{12}_{\al\be}\sim\de_{\al\be}O(z-x_a)+(1-\de_{\al\be})O(z-x_a)^2,
\eq
while the other matrix element are continuae in the marked points.
It can be checked that these matrices define the Lie subalgebra  
Lie$({\cal G})$ in the Lie algebra of smooth maps.

The form $\om'$ (\ref{4.30}) is invariant under the gauge transform
$$
 \bar{\cal A}\rar f^{-1}\bp f+ f^{-1}\bar{\cal A}f,
~~f\in{\cal G},
$$
\beq{gt}
{\cal A}\rar f^{-1}\ka\p f+ f^{-1}{\cal A}f
\eq
$$
g_a\rar g_af_a,~p_a\rar f_a^{-1}p_af_a,~f_a=f(x_a,\bar{x}_a).
$$

We identify the dual space $($Lie$)^*({\cal G})$ with matrices living
on $\Si_{g,n}$ by means of the bilinear form
$$
\int_{\Si_g}<<x,y>>,~(x\in{\rm Lie}({\cal G}),~y\in{\rm Lie}^*({\cal G}).
$$
Due to (\ref{ge2}) the matrix elements $y\in($Lie$)^*({\cal G})$ have 
the following form in the marked points
\beq{ds2}
y^{21}_{\al\al}\sim\sum_{k\geq 1}b_k\p^k \de(x_a),~~
y^{21}_{\al\be}\sim\sum_{k\geq 2}b_{k,\al,\be}\p^k \de(x_a),~(\al\neq\be).
\eq
\beq{dds2}
y^{JK}_{\al\be}\sim\sum_{k\geq 0}b_{k,\al,\be}\p^k \de(x_a),
~~(J\neq 2,~K\neq 1).
\eq

The form $\om'$ (\ref{4.30}) is degenerated on the orbits of the gauge group.
The condition (\ref{4.60a}) is similar to the partial gauge fixing. 
We choose ${\cal A}$ in the form
\beq{4.8}
{\cal A}=\mat{0}{E}{\ti{T}}{-2A},
\eq
where $\ti{T}$ is given by (\ref{4.3}). This form of ${\cal A}$ is the gauge
transform of (\ref{4.60a}) by
$$
f=\mat{{\bf 1}_N}{0}{A}{{\bf 1}_N}.
$$

As usual, the curvature tensor $F({\cal A},\bar{\cal A})$ is the moment map
$$
m:({\cal A},\bar{\cal A},{\bf p})\rar({\rm Lie})^*({\cal G}).
$$
The flatness (\ref{4.50}) means that we take $m=0$.
It allows to define $\bar{\cal A}$.
The general solution of (\ref{4.50}) depends on two fields $\bA$ and $\mu$
and takes the form
\beq{4.10}
   \bar{\cal A} =\mat{-\bA+\oh\p\mu-\ka^{-1}\mu A}{-\ka^{-1}\mu}
{-\ka^{-1}\mu\ti{T}-\ka\p(\bA-\oh\p\mu+\ka^{-1}\mu A)}
{-\bA+\ka^{-1}\mu A-\oh\p\mu},
\eq
where $\bA$ takes value in Lie$(\SL{N})$.
It can be checked immediately that (\ref{4.50}) becomes the identity for
the $\GL{2}$ matrix elements $(1,1)$ and $(1,2)$ and coincides with
(\ref{4.5}) and (\ref{4.6}) for $(2,1)$ and $(2,2)$ respectively. 

Now let us discuss the boundary condition (\ref{4.6a}),(\ref{4.6b}).
Since $m=0$ all coefficients in 
(\ref{ds2}) and (\ref{dds2}) of 
$F^{2,1}({\cal A},\bar{\cal A})$ and $F^{2,2}({\cal A},\bar{\cal A})$
vanish. The expression $\bp \ti{T}$ in the matrix element $(2,1)$ 
has the terms proportional to $\p \de(x_a)$. Their cancellation lead 
to (\ref{4.6b}). In the similar way, using (\ref{dds2})
 we obtain (\ref{4.6a}) from $\bp A$ in the matrix element $(2,2)$.$\Box$

The flatness of these connections is equivalent to the compatibility of the 
linear system
\beq{4.9a}
(\ka\p-{\cal A})\left(
\begarc
\psi \\
\ka\p\psi 
\enarr
\right )=0,
\eq
\beq{4.9b}
(\bp-\bar{\cal A})\left(
\begarc
\psi \\
\ka\p\psi 
\enarr
\right )=0
\eq
Substituting in this system ${\cal A}$ (\ref{4.8}) and $\bar{\cal A}$
(\ref{4.10}) we obtain the linear system (\ref{4.1}), (\ref{4.2})
in Proposition 3.1.

The reduced symplectic form is read off from (\ref{4.30}),(\ref{4.8})
and (\ref{4.10}). It coincides with $\om$ (\ref{4.7}).

\bigskip
\noindent
{\bf 3. Isomonodromic deformations.}\\
To obtain nontrivial dynamical systems with Hamiltonians
we modified our previous procedure. Instead of the 
gauge group ${\cal G}$ defined in (\ref{gt}) consider its parabolic subgroup
$$
{\cal G}^P=\left\{\Si_{g,n}\rar \mat{a}{0}{b}{a^{-1}}\otimes\SL{N}\right\},
$$
Due to this choice 
there are no moment constraints coming from the matrix elements $(2,1)$ of
$\GL{2}$.
Therefore, the conditions  (\ref{4.5}),(\ref{4.6b}) are absent in this case.
In this way we come to the manifold ${\cal K}_2^N$ with 
the constraints
(\ref{4.6}),(\ref{4.6a}) and the symplectic form (\ref{4.7}). Evidently
${\cal W}_2^N\subset{\cal K}_2^N\subset{\cal K}_2$.

Consider the following transformations of the fields in ${\cal K}_2^N$
\beq{4.11}
\mu\rar \mu+\xi,~~\bA\rar\bA+\f1{\ka}A\xi,~~T\rar T,~~ A\rar A,
\eq
where $\xi\in\Om^{(-1,1)}(\Si_g)$. 
It is the symmetry of the form $\om$ (\ref{4.7}).
The transformations (\ref{4.11}) are generated by the Hamiltonian
$$
{\cal H}_\xi=\int_{\Si_{g,n}}\xi(T-\frac{<A^2>}{N}).
$$
We put ${\cal H}_\xi=0$. It means, that the Sugawara relation
\beq{4.12a}
T=\f1{N}<A^2>
\eq
is the moment constraint with respect to the symmetry (\ref{4.11}).
We don't fix the gauge and thereby come to the space with  fields
$$
{\cal P}^N_2=\{A,\bA,{\bf p},\mu\}\sim{\cal K}_2^N/(T=\f1{N}<A^2>) .
$$
It is a bundle over the space of times $\{\mu\}$.
It follows  from (\ref{4.7}) that on this space $\om$ take  the form
\beq{4.13}
\oh\om=\int_{\Si_g}<\de A,\de\bA>
+2\pi i\sum_{a=1}^n\de<p_a,g_a^{-1}\de g_a>
-\f1{2\ka}\int_{\Si_g}\de<A^2>\de\mu.
\eq
Since we
did not fix the gauge of this transformation and preserve the field
$\mu$ the form $\om$ (\ref{4.13}) is degenerated on ${\cal P}^N_2$.
If one replaces
\beq{4.14a}
\bA'=\bA-\f1{\ka}\mu A,
\eq
 (\ref{4.13}) takes the canonical form
\beq{4.13b}
\oh\om=\int_{\Si_g}<\de A,\de\bA'>+2\pi i\sum_{a=1}^n\de<p_a,g_a^{-1}\de g_a>.
\eq

Now we pass to the finite-dimensional space of equivalent
complex structures
 ${\cal T}_{g,n}$ (the Teichm\"{u}ller space).
The tangent space to the Teichm\"{u}ller space is 
$H^1(\Si_g, \G)$. Note, that only elements
of  $H^1(\Si_g, \G)$ contribute in the second integral 
in (\ref{4.13}). According to the
Riemann-Roch theorem $H^1(\Si_g, \G)$ has dimension
\beq{dim2}
l_2=\dim H^1(\Si_g, \G)=3g-3+n.
\eq
We fix a reference point 
${\bf \mu}^0=(\mu^0_1,\ldots,\mu^0_{l_2})$.
Then
\beq{cst}
\mu=\sum_{s=1}^{l_2}t_s\mu_s^0.
\eq
 defines a new complex structure and $\bft=(t_1,\ldots,t_{{l_2}})$
are coordinates of the tangent vector to ${\bf \mu}^0$ in
 $H^1(\Si_g, \G)$.

 Expanding $\mu$ in the basis (\ref{cst}) we rewrite $\om$ as
\beq{4.14}
\oh\om=\om^0-\f1{\ka}\sum_s\de H_s\de t_s,~~
H_s=H_s(A,\bft)=\oh\int_{\Si_g}<A^2>\mu_s^0,
\eq
where
$$
\om^0=\int_{\Si_g}<\de A,\de\bA>+2\pi i\sum_{a=1}^n\de<p_a,g_a^{-1}\de g_a>,~~
(\bft=(t_1,\ldots,t_l),~\p_s=\p/\p t_s).
$$
We keep the same notations for
the space ${\cal P}^N_2$   
$$
\begin{array}{cc}
{\cal P}^N_2& \\
\downarrow& \sim{\cal R}=\{A,\bA,{\bf p}\}\\
{\cal T}_g &
\end{array}
$$
${\cal P}^N_2$ is the extended phase space.
The form $\om^0$ is nondegenerated on the fibers. In this situation
the equation of motion for any function $F$ on ${\cal P}^N_2$ takes the form
\cite{Ar}
$$
\ka\frac{d F}{d t_s}  =\ka\frac{\p F}{\p t_s}+\{H_s,F\}_{\om^0}.
$$
In addition, there are the consistency conditions for the Hamiltonians
({\em the Whitham equations})
\beq{WE}
\ka\p_sH_r-\ka\p_rH_s+\{H_r,H_s\}=0.
\eq
They are satisfied since the Hamiltonians (\ref{4.14}) commute.
On this stage the equations of motion are trivial.
\beq{4.18}
\p_sA=0,~~\p_s\bA=\oh A\mu_s^0~~(s=1,\ldots,l).
\eq
We call this system the {\em Hierarchy of the Isomonodromic Deformations}
(HID). This notion will be justified below.

The degenerated symplectic form (\ref{4.14}) is invariant under the
 gauge transformations \\
$\ti{\cal G}={\rm Map}(\Si_{g,n}\SL{N})$
\beq{4.17}
A\rar f^{-1}Af+f^{-1}\ka\p f,~\bA\rar f^{-1}\bA f+f^{-1}(\bp+\mu\p) f,~
p_a\rar f^{-1}(x_a)p_af(x_a),~g_a\rar g_af(x_a).
\eq
The flatness condition 
\beq{4.13a}
(\bp+\p\mu)A-\ka\p\bA+[\bA,A]=0.
\eq
 is the moment constraint generating this symmetry. 
It allows to consider the linear consistent system on $\Si_g$
\beq{4.19}
(\ka\p+A)\Psi=0,
\eq
\beq{4.20}
(\bp+\sum_st_s\mu^0_s\p+\bA)\Psi=0,
\eq
where $\Psi\in\Om^0(\Si_{g,n},V)$.
The monodromy matrix $Y\in{\rm Rep}(\pi_1(\Si_{g,n}))\to\SL{N}$ transforms
 solutions  as
$$
\Psi\rar\Psi Y.
$$
The isomonodromy of (\ref{4.19}),(\ref{4.20}) is the independence of $Y$
on the deformations of the complex structure of $\Si_{g,n}$.
\begin{predl}
 The linear system (\ref{4.19}),(\ref{4.20}) has the property of
 the isomonodromic deformations with respect to  the "times"  $t_s$ iff
the equations of motion (\ref{4.18}) are satisfied. In other words, 
(\ref{4.18}) are monodromy preserving equations.
\end{predl}
{\sl Proof}\\
It follows from (\ref{4.18}) that $\p_s$ commute with $(\ka\p+A)$ and 
$(\bp+\sum_st_s\mu^0_s\p+\bA)$. Thus, in addition to (\ref{4.19}),(\ref{4.20}) 
one has consistent equations
\beq{4.21}
\p_s\Psi=0,~~(s=1,\ldots,l).
\eq
Then it follows from (\ref{4.21}) that
\beq{4.22}
\p_sY=0,~~(s=1,\ldots,l).
\eq

Assume now that the monodromy of the linear system (\ref{4.19}),(\ref{4.20})
is the time independent (\ref{4.22}). Then (\ref{4.21}) is fulfilled. The first
equation of motion (\ref{4.18}) follows from (\ref{4.19}) and (\ref{4.21}) and
the second from (\ref{4.20}) and (\ref{4.21}). $\Box$

The equations (\ref{4.18}) are not very interesting, because they describe a free
motion. They become nontrivial after the Hamiltonian reduction with respect to
the gauge transformations (\ref{4.17}). The flatness (\ref{4.13}) is the moment 
equation. Consider the gauge fixing assuming that $\bar L$ parametrizes the orbits
 of the gauge group $\ti{\cal G}$ in the phase space ${\cal R}$
\beq{GF}
\bar{L}=f^{-1}\bA f+f^{-1}(\bp+\mu\p) f.
\eq
It allows partly to fix $f$.
At the same time the gauge transformation defines
$$
L=f^{-1}Af+f^{-1}\ka\p f.
$$
We keep the same notations for the transformed matrices $p_a$.
Substituting these two expressions in the moment equation we obtain
\beq{4.33}
(\bp+\p\mu)L-\ka\p\bar{L}+[\bar{L},L]=0,
\eq
where according to (\ref{4.0}),(\ref{4.6a}) Res$L|_{z=x_a}=p_a$.
The gauge fixing (\ref{GF}) and the moment constraint (\ref{4.33})
kill almost all degrees of freedom. The fibers  
${\cal R}^{red}=\{L,\bar{L},{\bf p}\}$ become finite-dimensional,
as well as the bundle
 ${\cal P}^{red,N}_2$. 
The form $\om$ (\ref{4.14}) on ${\cal P}^{red,N}_2$ is 
\beq{red}
\om^0=\int_{\Si_{g,n}}<\de L,\de\bar{L}>+
2\pi i\sum_{a=1}^n\de<p_a,g_a^{-1}\de g_a>,~~
H_s=H_s(L,\bft)=\oh\int_{\Si_{g,n}}<L^2>\mu_s^0.
\eq
But now the system is no long free because 
due to (\ref{4.33}) $L$ depends on $\bar{L}$ and ${\bf p}$.
The equations of motion (\ref{4.18}) take the form
\beq{4.34}
\ka\p_sL-\ka\p M_s+[M_s,L]=0,~~M_s=\p_s ff^{-1},
\eq
\beq{4.35}
\ka\p_s\bar{L}-(\bp+\mu\p)M_s+[M_s,\bar{L}]=L\mu^0_s.
\eq
The equation (\ref{4.34}) is the analog of the Lax equation. The essential 
difference is the differentiation with respect to the spectral parameter 
$\p$. The last equation determines the matrices $M_s$. 
These equations are nontrivial and for the genus $g=0,1$ reproduce 
the Schlesinger system, Elliptic Schlesinger system, multicomponent
generalization of the Painlev\'{e} VI equation \cite{LO}.
The equations (\ref{4.34}), (\ref{4.35}) along 
with (\ref{4.33}) are consistency conditions for the linear system 
\beq{4.36}
(\ka\p+L)\Psi=0,
\eq
\beq{4.37}
(\bp+\sum_st_s\mu_s^0\p+\bar{L})\Psi=0,
\eq
\beq{4.38}
(\p_s+M_s)\Psi=0,~~(s=1,\ldots,l_2).
\eq
As in Proposition 3.3 the equations (\ref{4.38}) provides the isomonodromy 
property of the system (\ref{4.36}), (\ref{4.37}) with respect to variations
 of the times $t_s$.

\bigskip
\noindent
{\bf 4.Scaling limit.}\\
Consider the limit $\ka\to 0$. The value $\ka=0$ is called critical.
The symplectic form $\om$ (\ref{4.14}) is singular in this limit.
Let us  replace  the times
$$
t_s\rar t_s^0+\ka t_s,
$$
and assume that the times $t_s^0, ~(s=1,\dots)$ are fixed. 
After this rescaling the form  (\ref{4.14}) become regular.
The rescaling procedure means that we blow up a vicinity
 of the fixed point $\mu_s^{(0)}$
 in ${\cal T}_{g,n}$ and the whole dynamic 
 is developed in this vicinity.
This fixed point is defined by the complex coordinates
\beq{fp}
w_0=z-\sum_st^0_s\ep_s(z,\bz),~~\bar{w}_0=\bz,~~
 (\p_{\bar{w}_0}=\bp+\sum_st_s^0\mu_s^0).
\eq
For $\ka=0$ the connection $A$ is transformed into the one-form $\Phi$ 
(the Higgs field)
$
\ka\p +A\rar \Phi,
$ 
(see (\ref{4.17})).
Let
$
L^0=\lim_{\ka\to 0}L,~~\bar{L}^0=\lim_{\ka\to 0}\bar{L}.
$
Then we obtain the autonomous Hamiltonian systems with the form
$$
\om^0=\int_{\Si_g}<\de L^0,\de\bar{L}^0>+
2\pi i\sum_{a=1}^n\de<p_a,g_a^{-1}\de g_a>
$$
and the commuting quadratic integrals (\ref{WE}). The phase space 
${\cal R}^{red}$ is the cotangent bundle to the 
moduli of stable holomorphic $\SL{N}$-bundles over $\Si_{g,n}$. These
systems are completely integrable for the $\SL{2}$ bundles \cite{Ne}.

The corresponding set of linear equations has the following form.
The level $\ka$ can be considered as the Planck constant (see 
(\ref{4.36})). We consider the quasi-classical regime
$$
\Psi=\phi\exp\frac{\cal S}{\ka},
$$
where $\phi$ is a group-valued function and ${\cal S}$ is a scalar phase.
Assume that
$$
\frac{\p}{\p\bar w_0}{\cal S}=0,~~
\frac{\p}{\p t_s}{\cal S}=0.
$$
In the quasi-classical limit we set
$$
 \p{\cal S}=\la.
$$
Then instead of (\ref{4.36}), (\ref{4.37}), (\ref{4.38}) we obtain
$$
(\la+L^0)\Psi=0,
$$
$$
(\bp_{{\bar w}_0}+\la\sum_st_s\mu_s^0+\bar{L}^0)\Psi=0,
$$
$$
(\p_s+M^0_s)Y=0,~~(s=1,\ldots,l_2).
$$
Note, that the consistency conditions for the first and the last equations
are the standard Lax equations (\ref{1.1}).

\section{${\cal W}_3$ and Isomonodromic deformations}
\setcounter{equation}{0}
\noindent
{\bf 1. ${\cal W}_3$-structure.}\\
Define the space $\ti{\cal W}_3$
of fields $(W,\rho),(T,\mu)$ on $\Si_{g,n}$,
where $(T,\mu)$ are the same as in the case of the projective structures,
$\rho$ is the $(-2,1)$-differential and a variation of $W$ is
$(3,0)$-differential \cite{F,GLM}.
We assume that behavior of $\rho$ near the marked points
 has the form (compare with (\ref{6.1}))
\beq{5.00}
\rho|_{z\rar x_a}\sim (t_{3,a,0}+t_{3,a,1}(z-x^0_a))\bp\chi_a(z,\bz).
\eq
The dual field $W$ has poles in the marked points
\beq{5.01}
W|_{z\rar x_a}\sim
\frac{W_{-3,a}}{(z-x_a)^3}+\frac{W_{-2,a}}{(z-x_a)^2}+\frac{W_{-1,a}}{(z-x_a)}
+\ldots
\eq

\begin{defi}
The space  of ${\cal W}_3$-structure on $\Si_{g,n}$ is the subset of fields
in $\ti{\cal W}_3$ that satisfy the equations
\beq{5.1}
\ka^2\p^3\mu-(\bp+\mu\p+2\p\mu)T=
\frac{2}{3}\ka^2\p^2(\p^2-\f1{\ka^2}T)\rho+\f1{\ka}\p(W\rho)-
\f1{\ka}W(\mu-\p\rho)
\eq
\beq{5.2}
(\bp+\rho\p^2+(\mu+2\p\rho)\p+3\p\mu+)W=
\ka^3\p[(\p^2-\f1{\ka^2}T)(\p\mu-\frac{2}{3}(\p^2-\f1{\ka^2}T)\rho)].
\eq
with fixed values of
\beq{5.2a}
{\bf T}_2=(T_{-2,1},\ldots,T_{-2,n})
\eq
 in (\ref{3.1b}) and 
\beq{5.2b}
{\bf W}_2=(W_{-3,1},\ldots,W_{-3,n})
\eq
in (\ref{5.01}).
\end{defi}

\begin{predl}
The equations (\ref{5.1}),(\ref{5.2}),
 are the compatibility conditions for the linear system
\beq{5.11a}
(\ka^3\p^3+\ka T\p+W)\Psi=0,
\eq
\beq{5.12a}
(\bp+\frac{2}{3}(\p^2-\f1{\ka^2}T)\rho-\p\mu+
(\mu-\p\rho)\p-\rho\p^2)\Psi=0,
\eq
where $\Psi$ is a section of $\Om^{(-1,0)}(\Si_{g,n})$.
\end{predl}

The space ${\cal W}_3$ can be endowed with
the symplectic form 
\beq{5.3}
\om=\int_{\Si_{g,n}}(\de T\de\mu+\de W\de\rho).
\eq
As for the projective connections the space ${\cal W}_3$ can be derived
from the space ${\cal N}_3$ of flat $\SL{3}$ connections
$$
{\cal N}_3=\{a,\bar{a}|F(a,\bar{a})=0\}
$$
over $\Si_{g,n}$ with the symplectic form (\ref{3.5}). In fact, (\ref{5.3})
is induced by (\ref{3.5}).
The flatness constraints generate the gauge symmetry of ${\cal N}_3$.
We assume that the matrix elements of the Lie algebra of the gauge group
are continuous in the marked points and, moreover,
\beq{5.3a}
x^{13}|_{z\rar x_a}=O(z-x_a)^2,~~x^{23}|_{z\rar x_a}=O(z-x_a).
\eq
(see (\ref{3.5a})). In fact, (\ref{5.3a}) defines a subalgebra in the Lie
algebra of continuous gauge transformations.

The partial gauge fixing with respect to the parabolic subgroup
\cite{F}
\beq{ps}
P=\mathr{*}{*}{0}{*}{*}{0}{*}{*}{*}
\eq
allows to pick up a special form of connections
\beq{5.4}
a=\mathr{0}{1}{0}{0}{0}{1}{W}{T}{0}
\eq
with the prescribed behavior of $W$ and $T$ in neighborhoods of the marked 
points. Then the exact form of the matrix $\bar{a}$ as well as (\ref{5.1}),
(\ref{5.2}) are extracted from the flatness condition.
The defining properties (\ref{5.3a}) of the gauge algebra allows to fix 
the coefficients of highest poles (\ref{5.2a}),(\ref{5.2b}).
They define  the value of the moment $m=F(a,\bar{a})$.
We generalize this procedure below.

The flatness 
is the compatibility conditions for the linear system
$$
(\ka\p-a)\left(
\begarc
\psi \\
\ka\p\psi \\
\ka^2\p^2\psi
\enarr
\right )=0,\\
~~(\bp-\bar{a})\left(
\begarc
\psi \\
\ka\p\psi \\
\ka^2\p^2\psi
\enarr
\right )=0,\\
$$
where $\psi\in\Om^{(-1,0)}(\Si_{g,n})$. For the special form of $a$ (\ref{5.4})
and $\bar{a}$ it leads to (\ref{5.11a}),(\ref{5.12a}) in Proposition 4.1.

\bigskip
\noindent
{\bf 2. ${\cal W}_3^N$ structures.}\\
As in {\bf 3.1} consider $\SL{N}$-bundle $V$ over $\Si_{g,n}$ with 
a fixed complex structure. Let $(A,\bA)$ be connections in $V$.
We generalize the ${\cal W}_3$ structure 
(\ref{5.1}),(\ref{5.2}) taking into account the gauge degrees of freedom.

Define the space 
$$
\ti{\cal W}_3^N=\ti{\cal W}_2^N\cup(W,\rho),
$$
where $\rho$ and $W$ satisfy (\ref{5.00}),(\ref{5.01}). 
Introduce the following matrices
\beq{5.5}
\ti{T}=T-3(A^2+\ka\p A),
\eq
\beq{5.6}
\ti{W}=W+TA-A^3-\ka(A\p A+\p A^2)-\ka^2\p^2A.
\eq
\begin{defi}
${\cal W}_3^N$ structure on $\Si_{g,n}$ is the subset of fields in
$\ti{\cal W}_3^N$
satisfying  the following identities:
\beq{5.7}
\bp \ti{W}-\ka\p f_5+\ti{W}f_1+\ti{T}f_3-3Af_5-f_7\ti{W}=0,
\eq
\beq{5.8}
\bp \ti{T}-\ka\p f_6+\ti{W}f_2+\ti{T}f_4-3Af_6-f_7\ti{T}=0,
\eq
\beq{5.9}
\bp A-\ka\p\bA+[\bA,A]=0.
\eq
\beq{5.9a}
A_{-1,a}=p_a,~(a=1,\ldots,n),
\eq
\beq{5.9b}
T_{-2,a}=\frac{3<(p_a^0)^2>}{N}
~(a=1,\ldots,n).
\eq
\beq{5.9c}
W_{-3,a}=\frac{<(p_a^0)^3>-3\ka<(p_a^0)^2>}{N},~(a=1,\ldots,n).
\eq
\end{defi}
The matrix coefficients $f_j$ have the form
\beq{f1}
f_1=-\frac{2}{3}(\p^2-\f1{\ka^2}T)\rho+\p\mu-\bA+\frac{2}{\ka}A\p\rho
-\f1{\ka}\mu A-\frac{2}{\ka^2}A^2\rho,
\eq
\beq{f2}
f_2=-\f1{\ka}(\mu-\p\rho)-\frac{3}{\ka^2}A\rho,
\eq
\beq{f3}
f_3=\ka\p f_1-\f1{\ka^2}\ti{W}\rho=
\eq
$$
-\frac{2}{3}\ka\p(\p^2-\f1{\ka^2}T)\rho+\ka\p^2\mu-\ka\p\bA+2\p(A\p\rho)
-\f1{\ka^2}\ti{W}\rho-\p(\mu A+\frac{2}{\ka}A^2\rho),
$$
\beq{f4}
f_4=\ka\p f_2+f_1-\f1{\ka^2}\ti{T}\rho=
\eq
$$
\f1{3}(\p^2-\f1{\ka^2}T)\rho-\frac{1}{\ka}A\p\rho-\bA+
\frac{1}{\ka^2}A^2\rho-\f1{\ka}\mu A,
$$
\beq{f5}
f_5=-\frac{\mu}{\ka}\ti{W}+\ka\p f_3=
\eq
$$
-\frac{2}{3}\ka^2\p^2(\p^2-\f1{\ka^2}T)\rho+\ka^2\p^3\mu-\ka^2\p^2\bA+
2\ka\p^2(A\p\rho)-\f1{\ka}\p(\ti{W}\rho)-\frac{\mu}{\ka}\ti{W}
-\ka\p^2(\mu A +\frac{2}{\ka}A^2\rho),
$$
\beq{f6}
f_6=-\frac{\mu}{\ka}\ti{T}+\ka\p f_4+f_3=
\eq
$$
-\frac{2}{3}\ka\p(\p^2-\f1{\ka^2}T)\rho+\ka\p^2\mu-\ka\p\bA+
\p(A\p\rho)-\f1{\ka^2}\ti{W}\rho-\frac{\mu}{\ka}\ti{T}
+\frac{3}{\ka}\p(A^2)\rho
-2\p(\mu A +\frac{2}{\ka}A^2\rho),
$$
\beq{f7}
f_7=\frac{3}{\ka}\mu A-\p\mu+f_4=
\eq
$$
\frac{2}{\ka}\mu A-\p\mu+\f1{3}(\p^2-\f1{\ka^2}T)\rho-
\frac{1}{\ka}A\p\rho-\bA+\frac{1}{\ka^2}A^2\rho.
$$
Note that (\ref{5.7}),(\ref{5.8}) is reduced to the standard
${\cal W}_3$ structure if $A=\bA=0$, while (\ref{5.9}) is the flatness
of the bundle $V$.

\begin{predl}
The equations (\ref{5.7}),(\ref{5.8}),
(\ref{5.9}) are the compatibility conditions for the linear system
\beq{5.11}
(\ka^3\p^3+3A\ka^2\p^2+\ka\ti{T}\p+\ti{W})\Psi=0,
\eq
\beq{5.12}
(\bp-f_1-f_2\ka\p-\rho\p^2)\Psi=0,
\eq
where  and $\Psi$ is a section of $\Om^{(-1,0)}(\Si_{g,n},{\rm Aut}V)$.
\end{predl}

Again ${\cal W}_3^N$ has a natural symplectic form
\beq{5.10}
\om=\int_{\Si_g}[-\frac{N}{\ka}\de T\de\mu-\frac{N}{\ka^2}\de W\de\rho+
3<\de A,\de\bA>]+6\pi i\sum_{a=1}^n\de<p_a,g_a^{-1}\de g_a>.
\eq

The proof of  Proposition and the derivation of $\om$ (\ref{5.10})
is based on the same procedure that
we already used in the ${\cal W}_2^N$ case. We consider 
 $\ka\p+{\cal A},~~\bp+\bar{\cal A}$ connections in the 
$(S(\GL{3}\otimes\GL{N}))$ vector bundle. The first component of 
${\cal A},\bar{\cal A}$ acts on the sections of 
$3$-jets of $\Om^{(-1,0)}(\Si_{g,n})$, while the second component acts
on the sections of the bundle $\Om^{(-1,0)}(\Si_{g,n},{\rm Aut}V)$.
We define the space ${\cal K}_3$ of connections $({\cal A},\bar{\cal A})$
and the set of coadjoint orbits (\ref{4.00}):
$$
({\cal A},\bar{\cal A},{\bf 1}_3\otimes{\bf p}).
$$
 We assume that ${\cal A}$ 
satisfies additional restrictions:\\
$\bullet$ the ${\cal A}$ component are gauge equivalent
to the special form
\beq{4.60}
{\cal A}\sim DS\otimes{\bf 1}_N+{\bf 1}_3\otimes(-A),
\eq
where
$$
DS=\mathr{0}{1}{0}{0}{0}{1}{W}{T}{0}
$$
$\bullet$ near the marked points $W$ has the form (\ref{5.01});\\
$\bullet$ near the marked points $T$ has the form (\ref{3.1a});\\
$\bullet$ near the marked points $A$ has the form (\ref{4.00}).

Proposition 4.1 follows from the following statement
\begin{predl}
${\cal W}_3^N$ is the subset of ${\cal K}_3$, satisfying the flatness condition
\beq{5.13}
\bp{\cal A}-\ka\p\bar{\cal A}+[{\cal A},\bar{\cal A}]=0.
\eq
\end{predl}
{\sl Proof.}\\
The symplectic form $\om'$ on ${\cal K}_3$ has the similar form  
as in the ${\cal K}_2$ case:
\beq{sf3}
\om'=\int_{\Si_{g,n}}<<\de{\cal A},\de\bar{\cal A}>>+
6\pi i\sum_{a=1}^n\de <p_a,g_a^{-1}\de g_a>
\eq
The gauge group symmetry of $\om'$ is
$$
{\cal G}=\{{\rm Map}(\Si_{g,n}\rar \SL{3N^2})\}.
$$
The action of ${\cal G}$ is the same as in (\ref{gt}).
For the matrix elements of Lie$({\cal G})$ we assume that they
are continuous in the marked points with additional restrictions
for the matrix elements $x^{13}_{\al\be}$,$x^{23}_{\al\be}$
$$
x^{13}_{\al\be}\sim\de_{\al\be}O(z-x_a)^2+(1-\de_{\al\be})O(z-x_a)^3,
$$
$$
x^{23}_{\al\be}\sim\de_{\al\be}O(z-x_a)+(1-\de_{\al\be})O(z-x_a)^2.
$$
Then for the dual space $y\in$Lie$^*({\cal G})$ we have
\beq{5.13b}
y^{31}_{\al\al}\sim\sum_{k\geq 2}b_k\p^k \de(x_a),~~
y^{31}_{\al\be}\sim\sum_{k\geq 3}b_{k,\al,\be}\p^k \de(x_a),~(\al\neq\be).
\eq
\beq{5.13a}
y^{32}_{\al\al}\sim\sum_{k\geq 1}b_k\p^k \de(x_a),~~
y^{32}_{\al\be}\sim\sum_{k\geq 2}b_{k,\al,\be}\p^k \de(x_a),~(\al\neq\be).
\eq
\beq{5.13c}
y^{JK}_{\al\be}\sim\sum_{k\geq 0}b_{k,\al,\be}\p^k \de(x_a),
~~(J\neq 3,~K\neq 1,2).
\eq
We choose ${\cal A}$ in the form
\beq{5.17}
{\cal A}=\mathr{0}{E}{0}{0}{0}{E}{\ti{W}}{\ti{T}}{-3A}
\eq
Substituting this form of ${\cal A}$ in (\ref{5.13}) we find $\bar{\cal A}$.
Solutions $\bar{\cal A}$ are 
parametrized by the fields $\mu,\rho$ and the matrix $\bA$. 
Then 
$$
\bar{\cal A}=
\mathr{f_1}{f_2}{-\rho/\ka^2}{f_3}{f_4}{-\mu/\ka}{f_5}{f_6}{f_7}.
$$
We set $f_1+f_4+f_7=-3\bA$. This condition along with algebraic equation for
the blocks $(J,K),~J=1,2,~K=1,2,3$ of (\ref{5.13}) allows to find 
$f_1,\ldots,f_7$ (\ref{f1})-(\ref{f7}).
 The differential identities
arising in the last row $J=3,K=1,2,3$ lead to 
(\ref{5.7}),(\ref{5.8}),(\ref{5.9}).
The behavior of the most singular terms near the
marked points (\ref{5.9a}),(\ref{5.9b}),(\ref{5.9c})
follows from the special form of Lie$^*({\cal G})$ (\ref{5.13a}),
(\ref{5.13b}),(\ref{5.13c}).

Moreover, $\om'$ (\ref{sf3}) gives $\om$ (\ref{5.10}) on ${\cal W}_3^N$
space.

Due to the special form of ${\cal A}$ and $\bar{\cal A}$
the compatible system  of differential equations
$$
(\ka\p-{\cal A})\left(
\begarc
\Psi \\
\ka\p\Psi \\
\ka^2\p^2\Psi
\enarr
\right )=0,
$$
$$
(\bp-\bar{\cal A})\left(
\begarc
\Psi \\
\ka\p\Psi \\
\ka^2\p^2\Psi
\enarr
\right )=0.
$$
 is equivalent to (\ref{5.11}),(\ref{5.12}). $\Box$

\bigskip
\noindent
{\bf 3.${\cal W}^N_3$ and isomonodromic deformations.}\\
To get a nontrivial phase flow we should get rid of from the restrictive
constraints (\ref{5.7}),(\ref{5.8}),(\ref{5.9b}) and (\ref{5.9c}).
The remaining constraints generate the gauge subgroup
$$
{\cal G}^P=\left\{{\rm Map}(\Si_{g,n}\rar P)\right\},
$$
where $P$ is the parabolic subgroup (\ref{ps}).
Thus, we come to the space ${\cal K}^N_3\subset{\cal K}_3$ of the 
fields\\
 $(W,\rho,T,\mu,A,\bA,p_a)$ with  constraints (\ref{5.9}),(\ref{5.9a}).

 The form $\om$ (\ref{5.10}) is invariant under the following
transformations
\beq{5.18}
A\rar A,~\bA\rar\bA+\f1{\ka}\xi_2A+\f1{\ka^2}\xi_3A^2,~
T\rar T,~\mu\rar\mu+\xi_2,~\rho\rar\rho+\xi_3,
\eq
where $\xi_j\in\Om^{(1-j,1)}(\Si_{g,n})$ with the same behavior near 
the marked points as $\mu$ and $\rho$.
The moment constraints generated by these transformations are
$$
T=\frac{3}{2N}<A^2>,~~
W=\f1{N}<A^3>.
$$
Substituting $T$ and $W$ in $\om$ (\ref{5.10}) we obtain
\beq{5.21}
\f1{3}\om=\int_{\Si_g}<\de A,\de\bA>+2\pi i\sum_{a=1}^n\de<p_a,g_a^{-1}\de g_a>
-\frac{1}{\ka}\int_{\Si_{g,n}}<\de A,A>\de\mu-
\frac{1}{\ka^2}\int_{\Si_{g,n}}<\de A,A^2>\de\rho.
\eq
Since we don't fix the gauge of the transformations (\ref{5.18}),
the form $\om$ becomes degenerated.
If one replace
$$
\bA\rar\bA'=\bA-\f1{\ka}\mu A-\f1{2\ka^2}\rho A^2
$$
$\om$ takes the canonical form
$$
\f1{3}\om=\int_{\Si_{g,n}}<\de A,\de\bA'>
+2\pi i\sum_{a=1}^n\de<p_a,g_a^{-1}\de g_a>.
$$
It is invariant under the gauge transformations
\beq{5.22}
A\rar fAf^{-1}+f\ka\p f^{-1},~~\bA'\rar f\bA'f^{-1}+f\bp f^{-1}.
\eq
The moment equation resulting from this symmetry is
$$
F(A,\bA'):=\bp A-\ka\p\bA+[\bA,A]=0,
$$
or in the original variables
\beq{5.24}
(\bp+\p\mu+\f1{2\ka}\p\rho A)A-\ka\p\bA+[\bA,A]=0.
\eq
In this case the constraints become nonlinear.

Now instead of infinite-dimensional space of smooth differentials
$\rho$ consider the finite dimensional space $H^1(\Si_{g,n},\G^2)$.
It has dimension
\beq{5.30a}
l_3=\dim H^1(\Si_{g,n},\G^2)=5g-5+2n.
\eq
Expanding $\rho$ in the basis of  $H^1(\Si_{g,n},\G^2)$ we obtain
$
\rho=\sum_{s=1}^{l_3}t_{3,s}\rho_s^0.
$
Then $\om$ (\ref{5.21}) gives
\beq{5.25}
\f1{3}\om=\om^0-\frac{1}{\ka}\sum_{s=1}^{l_2}\de H_{2,s}\de t_{2,s}
-\frac{1}{2\ka^2}\sum_{s=1}^{l_3}\de H_{3,s}\de t_{3,s}, 
\eq
$$
H_{2,s}=\f1{2}\int_{\Si_{g,n}}<A^2>\mu_s^{(0)},~
H_{3,s}=\f1{3}\int_{\Si_{g,n}}<A^3>\rho_s^{(0)},
$$
where
$$
~~\om^0=\int_{\Si_{g,n}}<\de A,\de\bA>
+2\pi i\sum_{a=1}^n\de<p_a,g_a^{-1}\de g_a>.
$$
The equations of motion defining by $\om$ are
\beq{5.28}
\p_rA=0,~~~(r=(k,s),~k=2,3,~~\p_r=\frac{\p}{\p t_{k,s}}),
\eq
\beq{5.29}
\p_r\bA'=0.
\eq
The solutions describe the free motion
\beq{5.30}
\bA=\bA_0+\f1{\ka}A_0\sum_st_{2,s}+\frac{1}{2\ka^2}A_0^2\sum_s t_{3,s}.
\eq

Making use of the gauge symmetry we represent the fields as
$$
L=fAf^{-1}+f\ka\p f^{-1},
$$
$$
\bar{L}'=f\bA'f^{-1}+f\bp f^{-1},
$$
where
$$
M_r=f^{-1}\p_rf.
$$
Thus we come to the finite-dimensional bundle 
$$
{\cal P}_3^{red,N}=\{L,\bar{L},{\bf p},\mu,\rho\}.
$$
The bundle ${\cal P}_3^N$ has the same fibers ${\cal R}^{red}$ as the 
bundle ${\cal P}_2^{red,N}$.

The equation (\ref{5.24}) takes the form
\beq{5.24a}
(\bp+\p\mu+\f1{2\ka}\p\rho L)L-\ka\p\bar{L}+[\bar{L},L]=0,
~~L|_{z\to x_a}=\frac{p_a}{z-x_a},
\eq
where
$$
\bar{L}=\bar{L}'+\f1{\ka}\mu L+\f1{2\ka^2}\rho L^2.
$$

The equation of motion (\ref{5.28}),
on the reduced phase space $(L,\bar{L}')$ takes the form of the Lax
equation
\beq{5.34}
\p_rL-\ka\p M_r+[M_r,L]=0.
\eq
The second equation (\ref{5.29}) allows to define the matrices $M_r$ in the Lax
equation
\beq{5.35}
\p_r\bar{L}'-\bp M_r+[M_r,\bar{L}']=0.
\eq

The equations (\ref{5.34}), (\ref{5.35}) with the flatness condition (\ref{5.24a})
are the compatibility conditions of the linear system
\beq{5.31}
(\ka\p+L)\psi=0,
\eq
\beq{5.32}
(\bp+\bar{L}')\psi=0,
\eq
\beq{5.33}
(\ka\p_r+M_r)\psi=0.
\eq
The last equation means that the monodromy matrix ${\cal M}$ of the linear
system (\ref{5.31}), (\ref{5.32}) on $\Si_{g,n}$ is independent with respect
to the ${\cal W}_3$ moduli $\mu_r,\rho_r$.

On the critical level $\ka=0$ the systems pass  into the Hitchin systems
with quadratic and cubic commuting integrals.  

Consider a rational curve $\Si_{0,n}$ with $n$ fixed marked points
$x^0_1,\ldots,x^0_n$. According to (\ref{dim2}) and (\ref{5.30a}) we have
$8n-8$ times. Since in this case $\bar L=0$, (\ref{5.24a}) takes the form
$$
[\bp+\p\sum_{a=1}^nt_{2,a}\bp \chi_a(z,\bz)
+\f1{2\ka}\p \sum_{a=1}^n (t_{3,a,0}+t_{3,a,1}(z-x^0_a))\bp\chi_a(z,\bz) L]L
=0.
$$
We remind that the times $t_{2,a}$ are related to the positions of the
marked points $(t_{2,a}=x_a-x_a^0)$. 
If the $t_3$ times vanish then solution of this equation
$$
L=\sum_{a=1}^n\frac{p_a}{w-x_a},~~(w=z-\sum_{a=1}^nt_{2,a}\chi_a(z,\bz))
$$
is the $L$ operator for the Schlesinger system.
For generic points $(t_{3,a}\neq 0)$ the 
equation is nonlinear and its solutions are unknown.

\section{Conclusion}
\setcounter{equation}{0}
The generalization of the previous analysis on the arbitrary ${\cal W}_k^N$
structures is straightforward, though the explicit calculations on 
intermediate steps become rather
long. In this section we  present some general formulas.

Let $W_j,j=2,\ldots,k$ be the set of $j$-differentials on $\Si_{g,n}$ and
$\rho_j,j=2,\ldots,k$ are the dual objects, $(W_2=T,\rho_2=\mu)$. In addition 
we have the gauge data $(A,\bA)$ and the coadjoint orbits in the
marked points $p_a,(a=1,\ldots,n)$
with the following behaviour
$$
W_j|_{z\rar x_a}\sim
\frac{W_{-j,a}}{(z-x_a)^j}+\frac{W_{-j+1,a}}{(z-x_a)^{j-1}}+
+\ldots,
$$
$$
\rho_j|_{z\rar x_a}\sim (t_{j,a,0}+t_{j,a,1}(z-x^0_a)+
\ldots+t_{j,a,j-2}(z-x_a)^{j-2})\bp\chi_a(z,\bz),
$$
and $A$ has poles of the first order (\ref{4.0}) with residues (\ref{4.6a}).
The highest order coefficients ${W_{-j,a}}$ are the linear combinations
of corresponding Casimirs up to the order $j$
$$
{W_{-j,a}}\sim\f1{N}(<(p_a^0)^j>+\ldots).
$$ 
Let  $\Psi$ be a section of 
$\Om^{(-\frac{k-1}{2},0)}(\Si_{g,n},{\rm Aut}V)$. Then
the ${\cal W}_k^N$ structure on $\Si_{g,n}$ is the  set of fields
$$
(W_j,\rho_j),~j=2,\ldots,k, (A,\bA),p_a,a=1,\ldots,n,
$$
providing the consistency of the following linear system
$$
(\ka^k\p^k+kA\ka^{k-1}\p^{k-1}+\ldots+\ti{W}_k)\Psi=0,
$$
$$
(\bp+\al_k\p^{k-1}+\ldots+\al_1)\Psi=0.
$$
Here 
$$
\ti{W}_k=W_k+AW_{k-1}+A^2W_{k-2}+\ldots+A^{k-2}W_2-A^k,
$$
and other coefficients can be read off from the flatness of the connections 
${\cal A},\bar{\cal A}$ in the $\SL{kN}$ bundle with
$$
{\cal A}=DS\otimes {\bf I}_N+{\bf I}_k\otimes (-A)
$$
$$
DS=\left (
\begin {array}{cccccc}
0     & 1      & 0     & \cdots   &       & 0 \\
 0    & 0      & 1     & \cdots   &       & 0 \\ 
\cdot & \cdot  & \cdot & \cdots   &       &\cdot \\
 
  0   & \cdot  & \cdot & \cdots  &  0     &  1\\                                                                                
 W_k   & W_{k-1}& \cdot & \cdots  & W_2    & 0
\end{array}
\right ).
$$
The symplectic form on ${\cal W}_k^N$ 
has the universal structure
\beq{7.1}
\om=\int_{\Si_{g,n}}[k<\de A,\de\bA>
-N\sum_{j=2}^k\f1{\ka^{j-1}}\de W_j\de \rho_j]
+2k\pi i\sum_{a=1}^n\de<p_a,g_a^{-1}\de g_a>.
\eq
We drop out details of calculations. Essentially they are the same as in
the case ${\cal W}_3^N$.

Again, the constraints on the fields 
$W_j,\rho_j,j=2,\ldots,k$ can be discarded by the restriction to the
parabolic subgroup of the gauge transformations.
The form acquire the following Abelian symmetry
$$
\rho_j\rar\rho_j+\xi_j,~(j=2,\ldots,k),~~
\bA\rar \bA+\sum_{j=2}^k\frac{A^{j-1}\xi_j}{\ka^{j-1}},
$$
where $\xi_j\in\Om^{(-j,1)}(\Si_{g,n})$.
This symmetry is generated by the constraints
$$
W_j=\frac{k<A^j>}{Nj}.
$$
Substituting this expression in (\ref{7.1}) we obtain the symplectic form
\beq{7.2}
\f1{k}\om=\int_{\Si_{g,n}}<\de A,\de\bA>+
2\pi i\sum_{a=1}^n\de<p_a,g_a^{-1}\de g_a>
-\sum_{j=2}^k\f1{\ka^{j-1}}\int_{\Si_{g,n}}<A^{j-1}\de A> \de \rho_j.
\eq
 Now we restrict the fields $\rho_j$ to the spaces 
$H^1(\Si_{g,n},\G^{j-1})$.
Let
$$
\rho_j=\sum_{s=1}^{l_j}t_{j,s}\rho_{j,s}^0, ~~~~
(l_j=\dim H^1(\Si_{g,n},\G^{j-1})=(2j-1)(g-1)+(j-1)n)
$$
be the expansion of $\rho_j$ in the basis of $H^1(\Si_{g,n},\G^{j-1})$.
Then we come to the following form \beq{7.3}
\f1{k}\om=\om^0-\sum_{j=2}^k\f1{\ka^{j-1}}\sum_{s=1}^{l_j}\de H_{j,s}\de t_{j,s},
\eq
where
$$
\om^0=\int_{\Si_{g,n}}<\de A,\de\bA>+2\pi i\sum_{a=1}^n\de<p_a,g_a^{-1}\de g_a>,
 ~~ H_{j,s}=\f1{j}\int_{\Si_{g,n}}<A^j>\rho_{j,s}^0.
$$
In terms of the field
$$
\bA'=\bA-\sum_{j=2}^k\f1{\ka^{j-1}}A^{j-1} \rho_j
$$
it takes the canonical form
$$
\f1{k}\om=\int_{\Si_g}[<\de A,\de\bA'>+2\pi i\sum_{a=1}^n\de<p_a,g_a^{-1}\de g_a>.
$$
It is a free system with solutions
$$
A=A_0=const, ~~
\bA({\bf t}=\bA_0+
\sum_{j=2}^k\frac{A^{j-1}}{\ka^{j-1}}
\sum_{s=1}^{l_j}t_{j,s}\rho_{j,s}^0, 
$$

After the Hamiltonians reduction with respect to the gauge symmetry (\ref{5.22})
we come to the extended phase space
$$
{\cal P}_k^N=\{L,\bar{L},{\bf p},\rho_j,j=2,\ldots,k\}.
$$
The equations of motion are the same as in the ${\cal P}_3^N$ case
(\ref{5.34}),(\ref{5.35}), where
$$
\bar{L}'=\bar{L}\sum_{j=2}^{k}\f1{\ka^{j-1}}\rho_{j}L^{j-1}.
$$

On the critical level we obtain the Hitchin systems with integrals
of order $j=2,\ldots,k$. In the case $k<N$ the number of integrals
is less then the dimension of the configuration space. For $k\geq N$ the
systems are completely integrable though the
integrals of order $j>N$ are not independent.
Thus, the distinguish case is $k=N$.

It is interesting to consider the limit $k\rar\infty$. 
Because the differential operators of an arbitrary order generate a
Lie algebra,
${\cal W}$ structure acquire the group-theoretical background as in 
the ${\cal W}_2$ case. As we argued above the simultaneous limit
$k\rar\infty,~N\rar\infty$ is distinguish and can lead to the
interesting field theories.

{\bf Acknowledgments}\\
{\sl The work of A.L. is supported in part by grants RFFI-98-01-00344
and 96-15-96455 for support of scientific schools. 
The work of M.O. is supported in part by grants RFFI-96-02-18046,
 INTAS 96-518 and 96-15-96455 for support of scientific schools. 
We are grateful to the Max-Planck-Institut f\"{u}r Mathemamatik in Bonn
for the hospitality, where this paper was prepared.
}

\small{

}
\end{document}